\documentclass[conference]{IEEEtran}

\usepackage{cite}
\usepackage{amsmath,amssymb,amsfonts}
\usepackage{amsthm}

\theoremstyle{definition}

\usepackage{algorithmic}
\usepackage{graphicx}
\usepackage{textcomp}
\usepackage{xcolor}
\usepackage{subcaption}
\usepackage{multirow}
\usepackage{booktabs}
\usepackage{listings}

\definecolor{codegreen}{rgb}{0,0.6,0}
\definecolor{codegray}{rgb}{0.5,0.5,0.5}
\definecolor{codepurple}{rgb}{0.58,0,0.82}
\definecolor{backcolour}{rgb}{0.95,0.95,0.92}

\lstdefinestyle{mystyle}{
    backgroundcolor=\color{backcolour},   
    commentstyle=\color{codegreen},
    keywordstyle=\color{magenta},
    numberstyle=\tiny\color{codegray},
    stringstyle=\color{codepurple},
    basicstyle=\ttfamily\footnotesize,
    breakatwhitespace=false,         
    breaklines=true,                 
    captionpos=b,                    
    keepspaces=true,                 
    showspaces=false,                
    showstringspaces=false,
    showtabs=false,                  
    tabsize=2
}

\def\BibTeX{{\rm B\kern-.05em{\sc i\kern-.025em b}\kern-.08em
    T\kern-.1667em\lower.7ex\hbox{E}\kern-.125emX}}
\begin{document}

\title{On Extending the Automatic Test Markup Language (ATML) for Machine Learning}

\author{Tyler Cody$^{\IEEEauthorrefmark{1} \IEEEauthorrefmark{2}}$,
        Bingtong Li$^{\IEEEauthorrefmark{1} \IEEEauthorrefmark{2}}$,
        Peter Beling$^{\IEEEauthorrefmark{1} \IEEEauthorrefmark{2}}$ \\
        $^{\IEEEauthorrefmark{1}}$Department of Industrial and Systems Engineering, Virginia Tech \\
        $^{\IEEEauthorrefmark{2}}$Responsible General Intelligence Lab, Virginia Tech \\
}

\maketitle

\begin{abstract}
This paper addresses the urgent need for messaging standards in the operational test and evaluation (T\&E) of machine learning (ML) applications, particularly in edge ML applications embedded in systems like robots, satellites, and unmanned vehicles. It examines the suitability of the IEEE Standard 1671 (IEEE Std 1671), known as the Automatic Test Markup Language (ATML), an XML-based standard originally developed for electronic systems, for ML application testing. The paper explores extending IEEE Std 1671 to encompass the unique challenges of ML applications, including the use of datasets and dependencies on software. Through modeling various tests such as adversarial robustness and drift detection, this paper offers a framework adaptable to specific applications, suggesting that minor modifications to ATML might suffice to address the novelties of ML. This paper differentiates ATML's focus on testing from other ML standards like Predictive Model Markup Language (PMML) or Open Neural Network Exchange (ONNX), which concentrate on ML model specification. We conclude that ATML is a promising tool for effective, near real-time operational T\&E of ML applications, an essential aspect of AI lifecycle management, safety, and governance.
\end{abstract}


\section{Introduction}

Operational test and evaluation (T\&E) is a critical aspect of systems engineering. There is a pressing need for messaging standards that support operational T\&E of machine learning (ML) applications \cite{king2019ai}. This need is especially exposed for edge ML--those applications which exist as embedded subsystems within micro-controllers or other hardware in deployed systems like industrial controls, robots, unmanned vehicles, satellites, and more \cite{munck2017test, guissouma2019virtual}. Operational T\&E of edge ML applications is needed to assess potential degradation due to lifecycle phenomena like concept drift \cite{gama2014survey, cody2022empirically} and adversarial attack \cite{chakraborty2021survey}. While there are calls in the literature for new test architectures for ML \cite{nishi2018test, cody2022test, cody2022systematic, cody2022combinatorial}, luckily, the automatic test community developed the IEEE Standard 1671 (IEEE Std 1671), named the Automatic Test Markup Language (ATML), for similar, albeit different, uses \cite{gorringe2011overview}. 

ATML is an XML-based standard used for exchanging Automatic Test System (ATS) data. ATML provides a format for describing, storing, and exchanging data about test and measurement systems \cite{seavey2005atml}. It was developed for general use \cite{jain2008implementing}, but is most widely applied to electronic systems like compute hardware and signal processors/generators. For this reason, a key extension of ATML was integration with IEEE Standard 1641 to include standards for sending signals (e.g., cosine waves) and receiving/handling related test results.

An important innovation of these signal-related extensions was to provide standards for defining the information within test signals. With ML applications, however, the information is rarely definable using a parameterized function like a sine or cosine wave, but rather requires transacting datasets between the test tool and the unit under test (UUT), i.e., the predictive model of the ML subsystem. In addition to this novelty, ML applications are often software-dominant. While ATML has been integrated with software architectures, that is different that using ATML to test software-dominant systems \cite{taylor2003incorporating}. These novelties beg the question of whether or not IEEE Std 1671 can be extended to address automatic test of ML applications, or if a new standard is needed all together.

In this paper, we explore the ability of IEEE Std 1671 to address the novelties of ML applications, including the integration of test data payloads and software-based tests. We model a variety of tests, including cross-validation and adversarial testing, using the XML schemas provided by the ATML standard. By providing this case study, researchers and practitioners are given a basis which can be adapted to suit their particular application. Our results suggest that only minor extensions of ATML may be necessary, depending on design decisions regarding the specification of test data, and that a new IEEE 1671.X standard may be sufficient to address them.

Importantly, while there are other standards and frameworks that are specifically designed for ML, like the Predictive Model Markup Language (PMML) \cite{guazzelli2009pmml} or Open Neural Network Exchange (ONNX) \cite{bai2019onnx}, they are focused on specifications of ML models themselves, not of the tests of those models. Similarly, recently developed systems theories of learning like abstract learning systems theory \cite{cody2023systems, cody2019systems, cody2020motivating, cody2022mesarovician, cody2022homomorphisms, cody2023cascading}, while perhaps motivating the semantics of related testing, are also not focused on the testing system itself. This can be said for other emerging standards like model cards \cite{mitchell2019model}. Therefore, this paper targets an important gap in standards for ML applications.

This paper is structured as follows. First, background is given on ATML. Then, ATML is applied to ML testing, along the way discussing the strengths and limitations of modeling choices. Finally, we end the paper with conclusions and a discussion of next steps. Ultimately, ATML is a promising standard for enabling effective and near real-time operational T\&E of ML applications---a critical aspect of artificial intelligence (AI) lifecycle management, safety, and governance.

\section{Background}

IEEE Standard 1671, also know as the ``Automatic Test Markup Language (ATML) for Exchanging Automatic Test Equipment and Test Information via XML,'' is part of a series of standards related to test and diagnostic systems developed by a working group within IEEE consisting of experts in the field of test equipment and systems. The goal of ATML is to enable the interchange of test information (including test requirements, test results, test procedures, test systems, and test environment descriptions) between various test development, test management, and product lifecycle management tools. ATML helps reduce the cost and complexity of integrating different test systems \cite{delgado2007atml}, as it simplifies the process of exchanging data between different components of an ATS. This makes it easier to reuse test software and to integrate third-party test systems \cite{wegener2004practical}.

ATML is made up of several parts, including:
\begin{itemize}
    \item \textbf{IEEE Std 1671}: Base Standard
    \item \textbf{IEEE Std 1671.1}: Test Description
    \item \textbf{IEEE Std 1671.2}: Instrument Description
    \item \textbf{IEEE Std 1671.3}: Unit Under Test (UUT) Description
    \item \textbf{IEEE Std 1671.4}: Test Configuration Description
    \item \textbf{IEEE Std 1671.5}: Test Adaptor Description
    \item \textbf{IEEE Std 1671.6}: Test Station Description
    \item \textbf{IEEE Std 1641}:  Signals and Test Definition
    \item \textbf{IEEE Std 1636.1}: Test Results
\end{itemize}
Each of these standards defines a set of XML schema documents, which can be used to validate ATML documents.

For a simple example, consider that we want to describe a simple test case for a digital multimeter unit (DMU) that measures the voltage across a resistor. In this case, we will focus on the Test Description part (IEEE Std 1671.1). The ATML Test Description may look like the code shown in Figure \ref{fig:dmu}. The XML structure represents a test description for a test case ``Measure Voltage'' which belongs to the test group ``Voltage Measurement''. The `TestRequirement' element describes that the voltage should be within a range from -10V to +10V. The `NumericLimitTestResult' element will store the result of the voltage measurement test once it is performed. The status of the test is initially set to ``NotTested''. Note, this is a simplified version and actual ATML documents may contain much more detailed information and can be complex.

\begin{figure}[t]
\begin{lstlisting}[language=XML]
<atml:TestDescription>
  <atml:TestGroup name="Voltage Measurement">
    <atml:Test name="Measure Voltage" id="Test1">
    
      <atml:TestRequirements>
        <atml:TestRequirement name="Voltage Range" requirementType="Range">
          <atml:TestLimit>
            <atml:Low>-10</atml:Low>
            <atml:High>+10</atml:High>
            <atml:Unit>V</atml:Unit>
          </atml:TestLimit>
        </atml:TestRequirement>
      </atml:TestRequirements>
      
      <atml:TestResult>
        <atml:NumericLimitTestResult name="Voltage Measurement" status="NotTested">
          <atml:TestLimit>
            <atml:Low>-10</atml:Low>
            <atml:High>+10</atml:High>
            <atml:Unit>V</atml:Unit>
          </atml:TestLimit>
        </atml:NumericLimitTestResult>
      </atml:TestResult>
      
    </atml:Test>
  </atml:TestGroup>
</atml:TestDescription>
\end{lstlisting}
\caption{Example test description for digital multimeter unit (DMU).}
\label{fig:dmu}
\end{figure}

\section{ATML for ML}

In the following, we consider how cross-validation, adversarial robustness testing, and concept drift tests can be modeled using ATML Test Descriptions.

\subsection{Cross-Validation}

To start exploring the application of ATML for ML applications, we can consider a test description for cross-validation. Cross-validation in ML typically involves splitting a dataset into a training set to train a model and a validation set to evaluate its performance. Repeating this process multiple times (folds) and averaging the results can help to obtain a robust performance estimation.

While ATML is primarily aimed at hardware testing rather than ML model validation, to describe an ML cross-validation test in the general structure of ATML, the approach shown in Figure \ref{fig:cv} could be taken. The XML structure gives a test description for a test case ``Cross Validation'' which belongs to the `TestGroup' element ``Machine Learning Model Validation''. In this case, the `TestRequirement' element describes the requirement for 5-fold cross-validation using the 'TestLimit' element. The `NumericLimitTestResult' element represents the validation score (e.g., accuracy) of the model, which we specify as wanting to be between 0.8 and 1.0 using the `TestLimit' element of the `NumericLimitTestResult' element by specifying 0.8 as the 'Low' limit and 1.0 as the 'High' limit. Similar to the DMU example, the status of the test is initially set to ``NotTested''.

\begin{figure}[t]
\begin{lstlisting}[language=XML]
<atml:TestDescription>
  <atml:TestGroup name="Machine Learning Model Validation">
    <atml:Test name="Cross Validation" id="Test1">
      
      <atml:TestRequirements>
        <atml:TestRequirement name="Number of Folds" requirementType="Count">
          <atml:TestLimit>
            <atml:Value>5</atml:Value>
          </atml:TestLimit>
        </atml:TestRequirement>
      </atml:TestRequirements>
      
      <atml:TestResult>
        <atml:NumericLimitTestResult name="Cross Validation Score" status="NotTested">
          <atml:TestLimit>
            <atml:Low>0.8</atml:Low>
            <atml:High>1.0</atml:High>
          </atml:TestLimit>
        </atml:NumericLimitTestResult>
      </atml:TestResult>
      
    </atml:Test>
  </atml:TestGroup>
</atml:TestDescription>
\end{lstlisting}
\caption{Test description for cross-validation.}
\label{fig:cv}
\end{figure}

Importantly, the code in Figure \ref{fig:cv} does not make reference to the test data. To indicate which dataset to use, we can assume that there is a separate system for providing data that the ATS system has access to, and that we will use ATML to specify how and when to use what data and how to configure related tests. Given that ATML does not natively support the concept of datasets used in ML, one way to do this is to leverage ATML's extensibility to define custom elements that refer to datasets.

For instance, a custom `TestRequirement' element to specify the name or identifier of the dataset to be used can be added, as shown in Figure \ref{fig:data}. In Figure \ref{fig:data}, a new `TestRequirement' element with the name ``Dataset ID'' is added, and the value ``DataSet\_123'' of the `TestLimit' element is the identifier for the dataset to be used for the test. This representation is still a simplification, as one may also need to represent more complex scenarios (like multiple datasets for multi-task learning). While adding custom elements may limit the interoperability of the ATML document, the use of identifiers allows the ATS to incorporate the use of test dataset payloads. 

\begin{figure}[t]
\begin{lstlisting}[language=XML]
<atml:TestDescription>
  <atml:TestGroup name="Machine Learning Model Validation">
    <atml:Test name="Cross Validation" id="Test1">
      
      <atml:TestRequirements>
        <atml:TestRequirement name="Number of Folds" requirementType="Count">
          <atml:TestLimit>
            <atml:Value>5</atml:Value>
          </atml:TestLimit>
        </atml:TestRequirement>

        <!-- Custom element to specify the dataset -->
        <atml:TestRequirement name="Dataset ID" requirementType="Identifier">
          <atml:TestLimit>
            <atml:Value>DataSet_123</atml:Value>
          </atml:TestLimit>
        </atml:TestRequirement>
      </atml:TestRequirements>
      
      <atml:TestResult>
        <atml:NumericLimitTestResult name="Cross Validation Score" status="NotTested">
          <atml:TestLimit>
            <atml:Low>0.8</atml:Low>
            <atml:High>1.0</atml:High>
          </atml:TestLimit>
        </atml:NumericLimitTestResult>
      </atml:TestResult>
      
    </atml:Test>
  </atml:TestGroup>
</atml:TestDescription>
\end{lstlisting}
\caption{Test description for cross-validation with reference dataset.}
\label{fig:data}
\end{figure}

\subsection{Adversarial Testing}
Adversarial robustness of an ML model can be tested by using adversarial examples, which are input sample modified in a way intended to cause the model to misclassify them. Figure \ref{fig:adv} shows how such test can be described in ATML. The ATML Test Description in Figure \ref{fig:adv} describes an adversarial robustness test where adversarial examples are created with a perturbation size defined by a `TestRequirement' epsilon (in this case, 0.1) and includes a reference to the dataset to be perturbed, similar to Figure \ref{fig:data}. The test case is named ``Adversarial Robustness Test'' using the `Test' element and belongs to the `TestGroup' element ``Machine Learning Model Validation'', as in the cross-validation example. The test result is represented by the robustness score, captured by a `NumericLimitTestResult' element within a `TestResult' element, which measures the proportion of adversarial examples that the model correctly classifies. We expect this score to be between 0.7 and 1.0, indicated by the `High' and `Low' values of the  `TestLimit' element.

\begin{figure}[t]
\begin{lstlisting}[language=XML]
<atml:TestDescription>
  <atml:TestGroup name="Machine Learning Model Validation">
    <atml:Test name="Adversarial Robustness Test" id="Test1">

      <atml:TestRequirements>
        <!-- Specify the dataset to be used -->
        <atml:TestRequirement name="Dataset ID" requirementType="Identifier">
          <atml:TestLimit>
            <atml:Value>DataSet_123</atml:Value>
          </atml:TestLimit>
        </atml:TestRequirement>

        <!-- Specify the epsilon value for adversarial perturbations -->
        <atml:TestRequirement name="Epsilon" requirementType="Float">
          <atml:TestLimit>
            <atml:Value>0.1</atml:Value>
          </atml:TestLimit>
        </atml:TestRequirement>
      </atml:TestRequirements>

      <atml:TestResult>
        <!-- Expected proportion of correctly classified adversarial examples -->
        <atml:NumericLimitTestResult name="Robustness Score" status="NotTested">
          <atml:TestLimit>
            <atml:Low>0.7</atml:Low>
            <atml:High>1.0</atml:High>
          </atml:TestLimit>
        </atml:NumericLimitTestResult>
      </atml:TestResult>
      
    </atml:Test>
  </atml:TestGroup>
</atml:TestDescription>
\end{lstlisting}
\caption{Test description for adversarial robustness test.}
\label{fig:adv}
\end{figure}

Adversarial robustness tests can be complex and involve multiple different types of adversarial attacks and various metrics. ATML can straighforwardly handle these cases, as shown in Figure \ref{fig:adv2}. In Figure \ref{fig:adv2}, we specify the epsilon value for the Fast Gradient Sign Method (FGSM) \cite{huang2017adversarial} adversarial attack and the corresponding robustness score. One would repeat the `TestRequirement' and `TestResult' blocks for each type of adversarial attack. Note, if a pre-defined order for the test is desired, a `Sequence' element can be added, as will be demonstrated in the following subsection.

\begin{figure}[t]
\begin{lstlisting}[language=XML]
<atml:TestDescription>
  <atml:TestGroup name="Machine Learning Model Validation">
    <atml:Test name="Adversarial Robustness Test" id="Test1">

      <atml:TestRequirements>
        <!-- Specify the dataset to be used -->
        <atml:TestRequirement name="Dataset ID" requirementType="Identifier">
          <atml:TestLimit>
            <atml:Value>DataSet_123</atml:Value>
          </atml:TestLimit>
        </atml:TestRequirement>

        <!-- Specify the epsilon value for Fast Gradient Sign Method (FGSM) adversarial perturbations -->
        <atml:TestRequirement name="FGSM Epsilon" requirementType="Float">
          <atml:TestLimit>
            <atml:Value>0.1</atml:Value>
          </atml:TestLimit>
        </atml:TestRequirement>
      </atml:TestRequirements>

      <atml:TestResult>
        <!-- Expected proportion of correctly classified FGSM adversarial examples -->
        <atml:NumericLimitTestResult name="FGSM Robustness Score" status="NotTested">
          <atml:TestLimit>
            <atml:Low>0.7</atml:Low>
            <atml:High>1.0</atml:High>
          </atml:TestLimit>
        </atml:NumericLimitTestResult>
      </atml:TestResult>

      <!-- Repeat TestRequirement and TestResult blocks for each different type of adversarial attack -->
      <!-- ... -->

    </atml:Test>
  </atml:TestGroup>
</atml:TestDescription>
\end{lstlisting}
\caption{Test description for multiple adversarial tests.}
\label{fig:adv2}
\end{figure}

\subsection{Drift Detection}

Detecting data drift in an ML model involves monitoring the statistical properties of the model's input data over time and raising an alert if a significant change (or ``drift'') is detected. To specify this type of test using ATML, the following considerations are essential:
\begin{enumerate}
    \item The test procedure, which would include steps for comparing the current data distribution against the reference data distribution.
    \item The parameters or threshold used to determine whether a signficant drift has occured.
    \item The expected outputs of the test, including any alerts or reports that should be generated if a drift is detected.
\end{enumerate}
These considerations are captured in the ATML Test Description shown in Figure \ref{fig:drift}. This ATML Test Description specifies a two-step test for data drift detection. Item (1) is captured in the `TestStep' element with value ``Step\_1'' and Items (2) and (3) are captured in the `TestStep' element with value ``Step\_2''. Their sequential order is specified by their inclusion within a `Sequence' element.

\begin{figure}[t]
\begin{lstlisting}[language=XML]
<atml:TestDescription>
  <atml:Test>
  
    <atml:Name>Data Drift Detection Test</atml:Name>
    <atml:UniqueIdentifier>DriftTest1</atml:UniqueIdentifier>
    <atml:Description>This test monitors the distribution of input data to the machine learning model and raises an alert if a significant drift is detected.</atml:Description>
    
    <atml:TestGroup>
      <atml:Name>Drift Detection</atml:Name>
      <atml:Sequence>
      
        <atml:TestStep id="Step_1">
          <atml:Name>Compare Current Data Distribution</atml:Name>
          <atml:Description>Compare the current distribution of input data against the reference distribution.</atml:Description>
        </atml:TestStep>
        
        <atml:TestStep id="Step_2">
          <atml:Name>Detect Significant Drift</atml:Name>
          <atml:Description>Raise an alert if a significant drift is detected based on predefined thresholds.</atml:Description>
        </atml:TestStep>
        
      </atml:Sequence>
    </atml:TestGroup>
    
  </atml:Test>
</atml:TestDescription>
\end{lstlisting}
\caption{Test description for drift detection with reference distribution.}
\label{fig:drift}
\end{figure}

The first step involves comparing the current distribution of input data against a reference distribution, and the second step involves detecting whether a significant drift has occurred based on predefined thresholds. The exact details of how the test is executed---how the distributions are compared, what thresholds are used, how the drift is detected, etc.---would need to be defined in the test execution environment or in a separate XML document. The ATML Test Description provides a high-level description of the test, but does not specify these implementation details.

If a closed form distribution is used as the reference distribution, for example, a Gaussian distribution, one could specify it in the ATML Test Description. ATML does not inherently include a mechanism for defining statistical distributions or mathematical formulas. These details would likely be handled in the test execution environment. However, one can specify that the method of drift detection is to compare the feature distributions against reference Gaussian distributions. Also, one can specify the expected mean and variance for these reference distributions as `Property' elements of the `Test' element or `TestStep' element, as shown in Figure \ref{fig:drift2}.

\begin{figure}[t]
\begin{lstlisting}[language=XML]
<atml:TestDescription>
  <atml:Test>
  
    <atml:Name>Data Drift Detection Test</atml:Name>
    <atml:UniqueIdentifier>DriftTest1</atml:UniqueIdentifier>
    <atml:Description>This test compares the distribution of each feature in the input data to a reference Gaussian distribution, and raises an alert if a significant drift is detected.</atml:Description>
    
    <atml:TestGroup>
      <atml:Name>Drift Detection</atml:Name>
      <atml:Properties>
      
        <atml:Property>
          <atml:Name>Reference Mean</atml:Name>
          <atml:Value>0</atml:Value>
          <atml:DataType>Double</atml:DataType>
        </atml:Property>
        
        <atml:Property>
          <atml:Name>Reference Variance</atml:Name>
          <atml:Value>1</atml:Value>
          <atml:DataType>Double</atml:DataType>
        </atml:Property>
        
      </atml:Properties>
      <atml:Sequence>
      
        <atml:TestStep id="Step_1">
          <atml:Name>Compare Feature Distributions</atml:Name>
          <atml:Description>Compare the distribution of each feature against a reference Gaussian distribution with the specified mean and variance.</atml:Description>
        </atml:TestStep>
        
        <atml:TestStep id="Step_2">
          <atml:Name>Detect Significant Drift</atml:Name>
          <atml:Description>Raise an alert if a significant drift is detected based on predefined thresholds.</atml:Description>
        </atml:TestStep>
        
      </atml:Sequence>
    </atml:TestGroup>
    
  </atml:Test>
</atml:TestDescription>
\end{lstlisting}
\caption{Test description for drift detection with reference Gaussian distribution.}
\label{fig:drift2}
\end{figure}

\section{Beyond Test Descriptions for ML}

ATML is a comprehensive suite and includes various schemas to represent different aspects of the testing process:
\begin{itemize}
    \item Test Description: This is what has been discussed so far. It covers the description of the test procedures, requirements, and expected results.
    \item Test Results: This schema represents the results of a test. It allows for capturing detailed results of individual test steps, as well as overall test results.
    \item Test Configuration: This schema is used to describe the configuration of the test system. This includes information about the equipment and software used for testing.
    \item Signal and Test Definition: This schema provides a way to describe signals, test conditions, and test responses. It includes detailed descriptions of analog and digital signals that can be used to describe complex sequences of signals.
    \item UUT (Unit Under Test) Description: This schema describes the unit that is being tested. This includes information about its components, characteristics, and physical configuration.
    \item Test Adapter Description: This schema is used to describe test adapters, which are hardware or software components that allow the test system to interface with the unit under test.
    \item Test Station Description: This schema is used to describe the test station itself. This includes information about the station's configuration, capabilities, and the equipment that it contains.
    \item Test Program Set Description: This schema is used to describe a set of tests that are intended to be run together as a group. It includes information about the order in which the tests should be run, the conditions under which each test should be performed, and the expected results.
\end{itemize}
In the following, we explore different schemas in the context of ML applications.

\subsection{Unit Under Test}

In the ATML context, a UUT typically refers to a physical hardware device or component. However, consider an ML model as the UUT. An example description is shown in Figure \ref{fig:uut}. It describes a convolutional neural network (CNN) for image classification as the UUT. To do this, it uses `UTTType', `UTTIdentifier', `UTTDescription', and `UTTCharteristics' elements. While it is apparently possible to do so with ATML, designers of ATS would likely be better served by incorporating PMML \cite{guazzelli2009pmml}, ONNX \cite{bai2019onnx}, or model cards \cite{mitchell2019model} as the basis for describing the UUT in detail. The PMML, ONNX, and model card descriptions could serve as a `UUTCharacteristic' element, and other `UTTCharacteric' elements could specify important details about the interfaces of the model, the software used to implement it and related tooling, and so forth. Similar to referencing datasets, references to unique identifiers within a database of PMML, ONNX, or model card descriptions could reduce the information content in the ATML messages. Ultimately, it is more important for the `UTT' element to include that information which supports the test execution environment, and, e.g., detailed model architecture descriptions like those specified using ONNX may be unnecessary for the purposes of test.

\begin{figure}[t]
\begin{lstlisting}[language=XML]
<atml:UUTDescription>
  <atml:UUT>
    <atml:UUTType>Machine Learning Model</atml:UUTType>
    <atml:UUTIdentifier>ID_123</atml:UUTIdentifier>
    <atml:UUTDescription>
      <atml:Software id="MLModel_1">
        <atml:Name>Convolutional Neural Network Model</atml:Name>
        <atml:Version>1.0.0</atml:Version>
        <atml:Description>This is a Convolutional Neural Network for image classification.</atml:Description>
      </atml:Software>
    </atml:UUTDescription>
    <atml:UUTCharacteristics>
      <atml:UUTCharacteristic name="Layers" value="10"/>
      <atml:UUTCharacteristic name="Input Shape" value="(32, 32, 3)"/>
      <atml:UUTCharacteristic name="Output Classes" value="10"/>
    </atml:UUTCharacteristics>
  </atml:UUT>
</atml:UUTDescription>
\end{lstlisting}
\caption{UUT description.}
\label{fig:uut}
\end{figure}

\subsection{Test Station Description}

In a traditional hardware testing environment, a test station refers to the specific set of hardware and software that is used to perform tests on a UUT. When testing ML models, the test station can be seen as the computing environment where the model is tested. This would include information about the hardware and software used to execute and test the model. An example ATML document for the Test Station Description is shown in Figure \ref{fig:tsd}.

\begin{figure}[t]
\begin{lstlisting}[language=XML]
<atml:TestStationDescription>
  <atml:TestStation>
  
    <atml:Name>ML Test Station 1</atml:Name>
    <atml:Description>Test station for machine learning models</atml:Description>
    <atml:TestStationType>Software</atml:TestStationType>

    <atml:Software id="Software_1">
      <atml:Name>Python</atml:Name>
      <atml:Version>3.8</atml:Version>
    </atml:Software>

    <atml:Software id="Software_2">
      <atml:Name>TensorFlow</atml:Name>
      <atml:Version>2.6</atml:Version>
    </atml:Software>

    <atml:Hardware id="Hardware_1">
      <atml:Name>NVIDIA GPU</atml:Name>
      <atml:Model>GeForce RTX 3090</atml:Model>
    </atml:Hardware>
    
  </atml:TestStation>
</atml:TestStationDescription>
\end{lstlisting}
\caption{Test station description.}
\label{fig:tsd}
\end{figure}

In this example, the test station includes Python as the programming language, TensorFlow as the ML library, and an NVDIA GeForce RTX 3090 GPU as the hardware used to run the tests. The software specifications of Python and TensorFlow are treated using `Software' elements of the `TestStation' element, while the compute hardware is treated using a `Hardware' element. This description helps ensure that the ML model is tested in a consistent environment, and that the test results can be reproduced if the tests are run again under the same conditions. It does not mandate that all test stations must use GeForce RTX 3090 GPUs, it simply notes that a GeForce RTX 3090 GPU is used at this test station. Remember, the ATML standard defines messages for the exchange of information within an ATS, it is not a systems modeling language like SysML meant to specify and enforce system requirements, rather, only requirements on the exchange of information within ATS.

\subsection{Test Adapter Description}

A test adapter in the ATML context is a hardware or software component that interfaces between the test station and UUT. In traditional hardware testing, a test adapter could be a physical devices that provides electrical or mechanical connections between the test station and the UUT. In ML, the concept of a test adapter may not directly map because ML applications deal primarily with software; however, test adapters could be interpreted in the ML context as a software layer that provides the interface between the test execution environment (test station) and the ML model (UUT). This might include data transformation functions, APIs, or other software utilities that facilitate interaction with the model.

An example is shown in Figure \ref{fig:tad}. In this brief example, the test adapter is a data preprocessing layer that uses the Scikit-Learn library and runs on an Intel Core i7 CPU. This preprocessing layer could handle tasks like scaling and normalizing input data, encoding categorical variables, or handling missing values. Similar to the ATML Test Station Description in Figure \ref{fig:tsd}, the Scikit-Learn library software is treated with a `Software' element within the `TestAdapter' element and the Intel Core i7 CPU is treated with a `Hardware' element.

\begin{figure}[t]
\begin{lstlisting}[language=XML]
<atml:TestAdapterDescription>
  <atml:TestAdapter>
  
    <atml:Name>Data Preprocessing Adapter</atml:Name>
    <atml:Description>This adapter handles the preprocessing of data for the machine learning model.</atml:Description>

    <atml:Software id="Software_1">
      <atml:Name>Scikit-Learn</atml:Name>
      <atml:Version>0.24</atml:Version>
    </atml:Software>

    <atml:Hardware id="Hardware_1">
      <atml:Name>CPU</atml:Name>
      <atml:Model>Intel Core i7</atml:Model>
    </atml:Hardware>
    
  </atml:TestAdapter>
</atml:TestAdapterDescription>
\end{lstlisting}
\caption{Test adapter description.}
\label{fig:tad}
\end{figure}

\subsection{Failure Reporting with Test Results Descriptions}

In the context of ATML, test failures or faults can be specified as a `TestResults' element. A test result in ATML can include information such as the identifier of the test, the timestamp of the test, the status of the test (pass or fail), and a diagnostic message in case of failure. To report a failure due to a low cross-validation score in an ML model test, the code shown in Figure \ref{fig:trd} could be used.

\begin{figure}[t]
\begin{lstlisting}[language=XML]
<atml:TestResultsDescription>
  <atml:TestResults>
    <atml:TestResult>
      <atml:Test>
        <atml:UniqueIdentifier>CVTest1</atml:UniqueIdentifier>
      </atml:Test>
      <atml:Timestamp>2023-06-23T10:00:00</atml:Timestamp>
      <atml:Status>Failed</atml:Status>
      <atml:Diagnostic>
        <atml:Message>Cross-validation score was below the accepted threshold.</atml:Message>
      </atml:Diagnostic>
    </atml:TestResult>
  </atml:TestResults>
</atml:TestResultsDescription>
\end{lstlisting}
\caption{Test results description for failure reporting.}
\label{fig:trd}
\end{figure}

In this example, the `TestResult' element provides information about a specific test. The `UniqueIdentifier' element refers to the identifier of the test (which should match the identifier specified in the Test Description document). The `Status' element indicates that the test failed, and the `Diagnostic' element includes a message explaining the reason for the failure. The time of failure is captured using a `TimeStamp' element and, in this example, the failure is associated with a low cross-validation score, described by a `Message' element within a `Diagnostic' element.

\subsection{Test Program Set}

In the ATML framework, a Test Program Set (TPS) is a collection of tests that are to be performed on a UUT. The Test Program Set Description (TPSD) language is used to describe the TPS. A TPSD document can be used to specify cross-validation, adversarial robustness, and drift detection tests that one wants to run on a given ML model (UUT). An example is shown in Figure \ref{fig:tpsd}. Within the `TestProgramSet' element, a `TestGroup' element is used to list the tests. The cross-validation, adversarial, and drift tests are described using `TestRef' elements. The `UniqueIdentifier' elements can link the `TestRef' elements to other ATML Test Description documents where the details are specified, e.g., those shown in Figures \ref{fig:data}, \ref{fig:adv}, and \ref{fig:drift}.

\begin{figure}[t]
\begin{lstlisting}[language=XML]
<atml:TestProgramSetDescription>
  <atml:TestProgramSet>
  
    <atml:Name>ML Model Test Program</atml:Name>
    <atml:UniqueIdentifier>MLTestProgram1</atml:UniqueIdentifier>
    <atml:Description>This test program includes a cross-validation test, an adversarial robustness test, and a drift detection test for a machine learning model.</atml:Description>
    
    <atml:TestGroup>
      <atml:Name>ML Model Tests</atml:Name>
      <atml:Tests>
      
        <atml:TestRef>
          <atml:UniqueIdentifier>CVTest1</atml:UniqueIdentifier>
        </atml:TestRef>
        
        <atml:TestRef>
          <atml:UniqueIdentifier>AdversarialTest1</atml:UniqueIdentifier>
        </atml:TestRef>
        
        <atml:TestRef>
          <atml:UniqueIdentifier>DriftTest1</atml:UniqueIdentifier>
        </atml:TestRef>
        
      </atml:Tests>
    </atml:TestGroup>
    
  </atml:TestProgramSet>
</atml:TestProgramSetDescription>
\end{lstlisting}
\caption{Test program set.}
\label{fig:tpsd}
\end{figure}

\subsection{Signal and Test Definition}

In addressing the use of datasets in ML testing, it is natural to consider, could the signal and test definition be used to describe datasets? Or is it better to use a reference indicator or identifier for the datasets? In ATML, the Signal Description Language (SDL) and the Test Description Language (TDL) are primarily intended for describing the characteristics of signals used in testing hardware devices, and the test steps and sequences, respectively.

While it is theoretically possible to use these languages to describe certain aspects of a dataset for an ML model, it may not be practical or efficient to do so. Datasets used in ML can be extremely complex and multidimensional, with diverse types of data (continuous, categorical, ordinal, etc.), varying levels of granularity, and various relationships between different data points. Additionally, SDL and TDL do not have built-in support for some important characteristics of ML datasets, such as the distribution of values, the balance of classes in classification problems, or the presence of missing or outlier values. 

In most cases, it would be more efficient and effective to refer to a dataset by a unique identifier or reference. Then, other tools or standards designed specifically for describing datasets, such as the Data Documentation Initiative (DDI) \cite{vardigan2008data} or the W3C's Data Catalog Vocabulary (DCAT) \cite{world2014data}, can be used to provide a detailed description of the dataset. When working with small, simple datasets, ATML's SDL and TDL languages could potentially be used. But for complex, real-world ML datasets, it would likely be more effective to use a combination of identifiers and tools/standards designed specifically for data description.

\section{Conclusion}

In this paper, we applied IEEE Std 1671 ATML to ML. We modeled test descriptions, a UUT, a test station description, a test adapter description, test results description, and test program set. We also discussed the limits of signal and test definitions for describing machine learning inputs and datasets. Our results suggest that with only minor extensions of ATML, many key information exchange concerns of an ATS for ML can be addressed. Depending on design decisions regarding the how datasets and models are specified, a new IEEE 1671.X standard may be sufficient to address the unique aspects of automatic testing for ML. 

In future work, we would like to implement an ATS using a combination of ATML and open-source ML software and hardware. It is crucial to practically demonstrate the use of ATML. Similarly, it is important to show how the ATML standards are different from and interact with other standards, like PMML, ONNX, etc. Ultimately, ATML should be used to support broader efforts to introduce open architectures into the way AI and autonomous systems are developed and operationalized.

\bibliographystyle{IEEEtran}
\bibliography{ref}

\end{document}